\documentclass[twocolumn,aps,amssymb,superscriptaddress]{revtex4-1}
\usepackage[final]{changes}
\usepackage[colorlinks=true,citecolor=blue,linkcolor=blue,urlcolor=blue]{hyperref}
\usepackage{amsmath}
\usepackage{amssymb}
\usepackage{bbm}
\usepackage{xcolor}
\usepackage{graphicx}
\usepackage{dcolumn}
\usepackage{bm}
\usepackage{hyperref}
\usepackage{afterpage}
\usepackage{tabularx}
\graphicspath{{Latexpicture/}}

\begin{document}

\title{Dissipation-assisted few-photon optical diode }

\author{Teng-Fei Xiao}
\affiliation{Hunan Key Laboratory for Micro-Nano Energy Materials and Devices\\ and School of Physics and Optoelectronics, Xiangtan University, Hunan 411105, China}

\author{Junlong Tian}
\affiliation{Department of Electronic Science, College of Big Data and Information Engineering, Guizhou University, Guiyang 550025, China}
\author{Jie Peng}
\email{jpeng@xtu.edu.cn}
\affiliation{Hunan Key Laboratory for Micro-Nano Energy Materials and Devices\\ and School of Physics and Optoelectronics, Xiangtan University, Hunan 411105, China}
\begin{abstract}
We studied the coherent transport of one or two photons in a one-dimensional
waveguide chirally coupled to a dissipative nonlinear cavity. The scattering amplitudes were derived analytically. With the assist of dissipation, we can realize an ideal
optical diode at the single-photon level, i. e., the transmittance is unity from one side and zero form the other side. The working area and properties of the two-photon diode are also found. This work details the relation between the diode effect and our system parameters, especially dissipation, which may find it applications in nonreciprocal quantum devices and quantum networks.

\end{abstract}


\date{\today}

\maketitle

\section{Introduction}\label{I}

One of the most promising prospects for future information and communication
technologies relies on quantum networks \cite{1} with integrated
optoelectronic components, some of which have been proposed and studied
\cite{2,3}. In these integrated networks, optical diodes that allow for
unidirectional signal propagation are essential. Optical diodes, also known as
optical isolators \cite{4}, require the ability to break the Lorentz reciprocity
\cite{5}. In the past decade, the non-reciprocal propagation of light has been
extensively studied through various physical mechanisms \cite{6,7,8,9,10}.
Recently, non-reciprocal quantum effects have been theoretically explored,
including non-reciprocal photon blockade \cite{11,12} and non-reciprocal
quantum entanglement \cite{13,14}. Some experiments have also reported
unidirectional propagation of light in metamaterials \cite{15,16,17,18} and
polarization-dependent light transport (chirality) in nanofibers \cite{19,20}.

Meanwhile, realizing optical diodes at few-photon level is important since
single photons are ideal carriers in quantum information processing. At the
same time, to ensure the normal propagation of optical information, parasitic
reflections between optical devices must be completely suppressed at the
single-photon level, as they have detrimental effects on optical devices based
on interference designs. Recently, single-photon diodes have been successfully
realized, in \cite{21,22,23,24,25}. Among them \cite{21}, a photon blockade
effect was utilized to achieve a photodiode at the level of two photons. In
chiral quantum optics \cite{26,27,28,29,30,31,32,33,34,zhihai,zheng}, light propagating in
opposite directions are coupled to emitters with different strengths. This
asymmetric coupling characteristic enables the system to achieve
unidirectional light transmission. Taking use of this technology, a single-photon router is realized in \cite{35}, and a two-photon
diode is implemented in \cite{36,37}. However, the vital role of dissipation can play in implementing the diode is not emphasized.

In this paper, we propose a dissipation-assisted few-photon diode based on the chiral coupling
between a one-dimensional waveguide and a dissipative Kerr-type nonlinear
cavity. We solve both the one- and two-photon scattering problem
analytically by the real space method. The two-photon correlation has been studied in such system \cite{liao1}, where the dissipation and chirality are not considered. The right- and left- going photons are
coupled to the nonlinear cavity with chiral coupling $\gamma_{1}$ and
$\gamma_{2}$ respectively. When the cavity dissipation $\kappa=\pm(\gamma_1-\gamma_2)>0$ under the resonance condition, the right- or left- going photons are blocked, respectively, so that a single photon diode is realized. Furthermore, if $\gamma_2=0$ ($\gamma_1=0$), then the transmittance of the left-going (right-going) photons reach unity, so that an ideal diode is realized. The two-photon transmission amplitudes consist of a plane wave part and a bound state part, which vanishes at specific positions and parameters. We obtain the working area and properties of the two-photon diode related to system parameters. Our work shows dissipation is a necessary condition for the single-photon diode in our system. Meanwhile, a thorough study on the relation between the few-photon diode and system parameters enables practical designs of such nonreciprocal devices in quantum networks.

This paper is organized as follows. In Sec. \textcolor{blue}{II}, we introduce the physical model.
Then we study the one-photon and two-photon
scattering in Secs. \textcolor{blue}{III} and \textcolor{blue}{IV} respectively. First we derive the analytical solution of scattering amplitudes and transmittance. Then we obtain the working area and properties of the photon diode related with system parameters. Finally,
we conclude in Sec. \textcolor{blue}{V}.

\section{MODEL AND HAMILTONIAN}\label{II}

The physical model consists of an infinitely long one-dimensional waveguide
and a nonlinear cavity located at the origin. The chiral coupling between a
nonlinear cavity and a waveguide, enables photons to tunnel between the
waveguide and the nonlinear cavity. By linearizing \cite{38,39} the dispersion
of the waveguide photon and considering the effect of environment, we obtain
an effective Hamiltonian ( $\hbar=1$ )

\begin{figure}[t]
\vspace{-30.0 mm}
\par
\begin{center}
\hspace{-2.5cm} \includegraphics[scale=0.3]{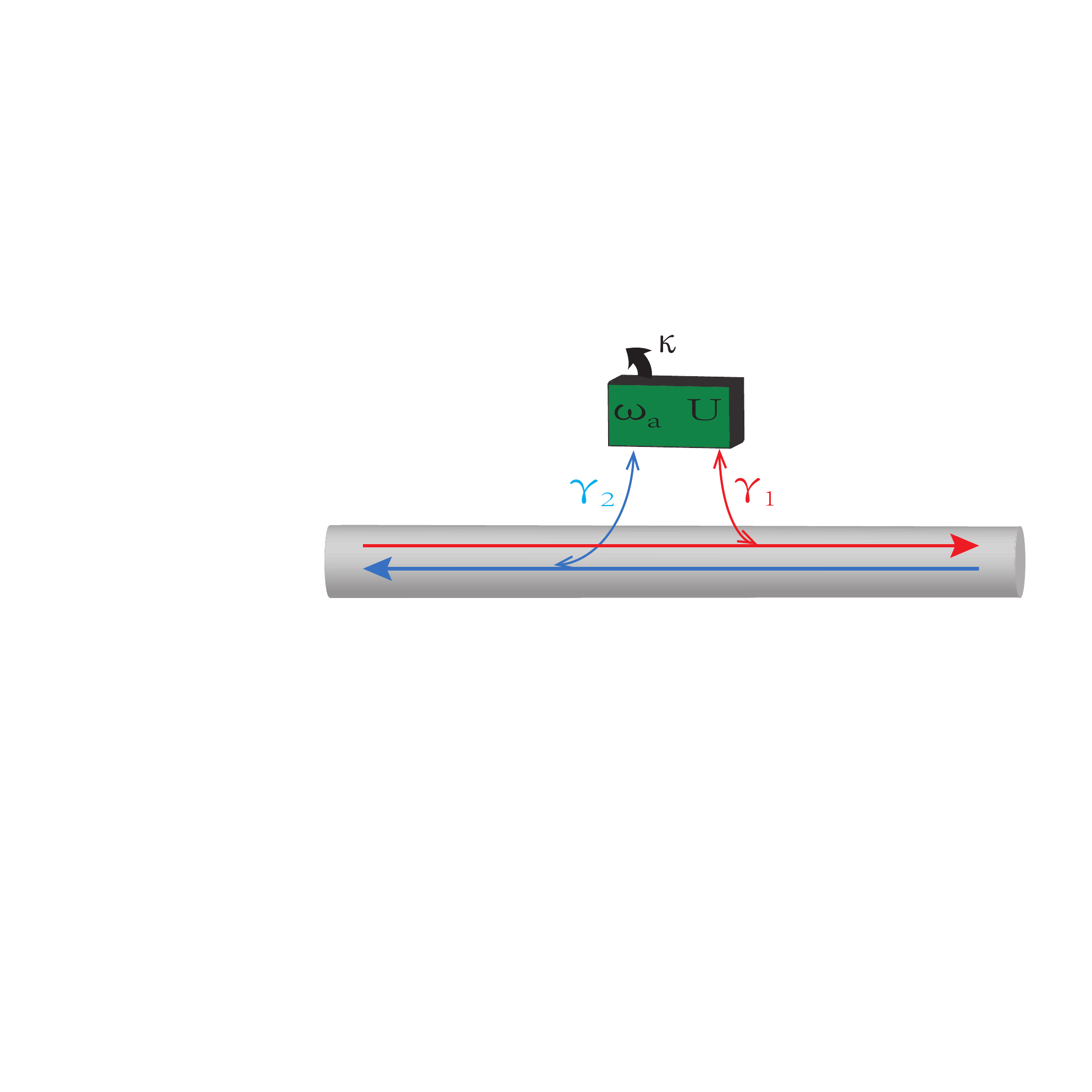}
\end{center}
\par
\renewcommand\figurename{\textbf{FIG.}}
\vspace{-40.0 mm}\caption{Schematic diagram of a 1D waveguide coupled to a
nonlinear cavity located at the origin. The photons incident from the left
side of the waveguide are scattered after passing through the nonlinear
cavity.}%
\label{Fig1}%
\end{figure}%
\begin{align}
\hat{H} &  =\left(  \omega_{a}-i\frac{\kappa}{2}\right)  \hat{a}^{\dag}\hat
{a}+U\hat{a}^{\dag}\hat{a}^{\dag}\hat{a}\hat{a}\nonumber\\
&  -i\upsilon_{c}\int dx\left[  \hat{r}^{\dag}\left(  x\right)  \frac
{\partial}{\partial x}\hat{r}\left(  x\right)  -\hat{l}^{\dag}\left(
x\right)  \frac{\partial}{\partial x}\hat{l}\left(  x\right)  \right]
\nonumber\\
&  +\sqrt{\upsilon_{c}\gamma_{1}}\int dx\delta\left(  x\right)  \left(
\hat{r}^{\dag}\left(  x\right)  \hat{a}+\hat{r}\left(  x\right)  \hat
{a}^{\dagger}\right)  \nonumber\label{1}\\
&  +\sqrt{\upsilon_{c}\gamma_{2}}\int dx\delta\left(  x\right)  \left(
\hat{l}^{\dag}\left(  x\right)  \hat{a}+\hat{l}\left(  x\right)  \hat
{a}^{\dagger}\right)  .
\end{align}
Here, $\hat{a}$ and $\hat{a}^{\dag}$ are the annihilation and creation
operators of the cavity mode with frequency $\omega_{a}$, and dissipation rate
$\kappa$. The second term describes the nonlinear interaction from the Kerr
medium in the cavity, with strength $U$. In the real space, the Hamiltonian of
the free fields propagating in the waveguide is denoted by the third term,
where $\hat{r}^{\dag}\left(  x\right)  \left[  \hat{l}^{\dag}\left(  x\right)
\right]  $ is the bosonic operator creating a right-going (left-going) photon
with group velocity $\upsilon_{c}$ at position $x$. These operators fulfill
the commutation relations
\begin{align}
\left[  \hat{l}\left(  x\right)  ,\hat{l}^{\dag}\left(  x^{\prime}\right)
\right]   &  =\left[  \hat{r}\left(  x\right)  ,\hat{r}^{\dag}\left(
x^{\prime}\right)  \right]  =\delta\left(  x-x^{\prime}\right)  ,\label{2}\\
\left[  \hat{l}\left(  x\right)  ,\hat{r}^{\dag}\left(  x^{\prime}\right)
\right]   &  =0.
\end{align}
The last two terms in Eq. (\ref{1}) represent the chiral coupling between the
cavity and the waveguide, from the right-going and left-going photon with strength $\sqrt{\upsilon_{c}\gamma_{1}}$ and $\sqrt{\upsilon
_{c}\gamma_{2}}$ respectively. For simplicity, we define the even- and
odd-parity modes operators of the waveguide as
\begin{align}
\hat{c}_{e}^{\dag}\left(  x\right)    & =\frac{\sqrt{\gamma_{1}}}{\sqrt
{\Gamma}}\hat{r}^{\dag}\left(  x\right)  +\frac{\sqrt{\gamma_{2}}}%
{\sqrt{\Gamma}}\hat{l}^{\dag}\left(  -x\right)  ,\label{3}\\
\hat{c}_{o}^{\dag}\left(  x\right)    & =\frac{\sqrt{\gamma_{2}}}{\sqrt
{\Gamma}}\hat{r}^{\dag}\left(  x\right)  -\frac{\sqrt{\gamma_{1}}}%
{\sqrt{\Gamma}}\hat{l}^{\dag}\left(  -x\right)  .
\end{align}
Hence, the Hamiltonian Eq. (\ref{1}) can be rewritten as
\begin{align}
\hat{H} &  =\left(  \omega_{a}-i\frac{\kappa}{2}\right)  \hat{a}^{\dag}\hat
{a}+U\hat{a}^{\dag}\hat{a}^{\dag}\hat{a}\hat{a}\nonumber\\
&  -i\upsilon_{c}\int dx\left[  \hat{c}_{e}^{\dag}\left(  x\right)
\frac{\partial}{\partial x}\hat{c}_{e}\left(  x\right)  +\hat{c}_{o}^{\dag
}\left(  x\right)  \frac{\partial}{\partial x}\hat{c}_{o}\left(  x\right)
\right]  \nonumber\\
&  +\int dx\sqrt{\upsilon_{c}\Gamma}\delta\left(  x\right)  \left[  \hat
{c}_{e}^{\dag}\left(  x\right)  \hat{a}+\hat{c}_{e}\left(  x\right)  \hat
{a}^{\dag}\right]  .\label{4}%
\end{align}
Here, $\Gamma=\gamma_{1}+\gamma_{2}$ is introduced. We notice that the
interaction involves solely even modes, and photons in the odd modes evolve
freely within the waveguide. Hence, we will focus on the calculation of the
transport properties of the photons in even modes with
\begin{align}
\hat{H}^{\left(  e\right)  } &  =\left(  \omega_{a}-i\frac{\kappa}{2}\right)
\hat{a}^{\dag}\hat{a}+U\hat{a}^{\dag}\hat{a}^{\dag}\hat{a}\hat{a}\nonumber\\
&  -i\upsilon_{c}\int dx\hat{c}_{e}^{\dag}\left(  x\right)  \frac{\partial
}{\partial x}\hat{c}_{e}\left(  x\right)  \nonumber\\
&  +\int dx\sqrt{\upsilon_{c}\Gamma}\delta\left(  x\right)  \left[  \hat
{c}_{e}^{\dag}\left(  x\right)  \hat{a}+\hat{c}_{e}\left(  x\right)  \hat
{a}^{\dag}\right]  .\label{5}%
\end{align}
For notational simplicity, we set $\upsilon_{c}=1$ in the following.

\section{Single-Photon Scattering And Its Diode Effect}\label{III}

We first consider the single-photon scattering, where we have found a
single-photon diode effect. We suppose there is an incident photon from the
left of the waveguide (right-going), with momentum $k$. The wave function of the single
photon before scattering (incident state) is
\begin{equation}
\left\vert \psi\right\rangle =\int dx\phi_{k}\left(  x\right)  \hat{r}^{\dag
}\left(  x\right)  \left\vert \oslash\right\rangle ,\label{6}%
\end{equation}
where $\phi_{k}\left(  x\right)  =e^{\mathrm{i}kx}/\sqrt{2\pi}$. The
right-going mode is decomposed into the even and odd modes as $\hat{r}^{\dag
}\left(  x\right)  =\left[  \sqrt{\gamma_{1}}\hat{c}_{e}^{\dag}\left(
x\right)  +\sqrt{\gamma_{2}}\hat{c}_{o}^{\dag}\left(  x\right)  \right]
/\sqrt{\Gamma}$. $\left\vert \varnothing\right\rangle =\left\vert
0\right\rangle _{c}\left\vert \emptyset\right\rangle $ represents the state
where there is no photon in the cavity and waveguide. The
incoming state Eq. (\ref{6}) can thus be rewritten as
\begin{align}
\left\vert \psi\right\rangle  & =\frac{\sqrt{\gamma_{1}}}{\sqrt{\Gamma}}\int
dx\frac{1}{\sqrt{2\pi}}e^{\mathrm{i}kx}\hat{c}_{e}^{\dag}\left(  x\right)
\left\vert \oslash\right\rangle \nonumber\label{7}\\
& +\frac{\sqrt{\gamma_{2}}}{\sqrt{\Gamma}}\int dx\frac{1}{\sqrt{2\pi}%
}e^{\mathrm{i}kx}\hat{c}_{o}^{\dag}\left(  x\right)  \left\vert \oslash
\right\rangle .
\end{align}
Now we study the scattering problem within the spaces spanned by the even and
odd modes. The general one-photon scattering state of the system assumes the
following form:
\begin{align}
\left\vert \psi\right\rangle  &  =\frac{\sqrt{\gamma_{1}}}{\sqrt{\Gamma}%
}\left\vert \psi_{e}\right\rangle +\frac{\sqrt{\gamma_{2}}}{\sqrt{\Gamma}%
}\left\vert \psi_{o}\right\rangle ,\label{8}\\
\left\vert \psi_{e}\right\rangle  &  =\int dx\phi_{e}\left(  x\right)  \hat
{c}_{e}^{\dag}\left(  x\right)  \left\vert \varnothing\right\rangle +\phi
_{a}\hat{a}^{\dag}\left\vert \varnothing\right\rangle ,\label{8.a}\\
\left\vert \psi_{o}\right\rangle  &  =\int dx\phi_{o}\left(  x\right)  \hat
{c}_{o}^{\dag}\left(  x\right)  \left\vert \varnothing\right\rangle
.\label{8.b}%
\end{align}
The stationary Schr\"{o}dinger equation gives rise to the coupled equations
for the amplitudes
\begin{align}
0 &  =\left(  -i\frac{\partial}{\partial x}-\omega\right)  \phi_{e}\left(
x\right)  +\sqrt{\Gamma}\delta\left(  x\right)  \phi_{a},\label{9a}\\
0 &  =\sqrt{\Gamma}\phi_{e}\left(  0\right)  +\left(  \omega_{a}-i\frac
{\kappa}{2}-\omega\right)  \phi_{a}.\label{9b}%
\end{align}
Substituting the scattering ansaz $\phi_{e}\left(  x\right)  =e^{\mathrm{i}%
kx}\left[  \theta\left(  -x\right)  +t_{k}\theta\left(  x\right)  \right]  $
(where $\theta\left(  x\right)  $ is the step function) into Eqs. (\ref{9a})
and (\ref{9b}), we obtain
\begin{equation}
t_{k}=\frac{\omega_{k}-\omega_{a}+i\frac{\kappa-\Gamma}{2}}{\omega_{k}%
-\omega_{a}+i\frac{\kappa+\Gamma}{2}}.\label{10}%
\end{equation}
Transforming $\hat{c}_{e}^{\dag}\left(  x\right)  $ and $\hat{c}_{o}^{\dag
}\left(  x\right)  $ back to the right- and left-going operators, we obtain
scattering amplitudes
\begin{align}
\bar{t}_{k} &  =\frac{\omega_{k}-\omega_{a}+i\frac{\kappa-\left(  \gamma
_{1}-\gamma_{2}\right)  }{2}}{\omega_{k}-\omega_{a}+i\frac{\kappa+\Gamma}{2}%
},\label{11a}\\
\bar{r}_{k} &  =\frac{-i\sqrt{\gamma_{1}\gamma_{2}}}{\omega_{k}-\omega
_{a}+i\frac{\kappa+\Gamma}{2}}.\label{11b}%
\end{align}

When a single photon enters the cavity from the right of the waveguide (left-going), the
incident state reads
\begin{equation}
\left\vert \psi\right\rangle =\int dx\phi_{k}\left(  -x\right)  \hat{l}^{\dag
}\left(  x\right)  \left\vert \oslash\right\rangle . \label{12}%
\end{equation}
One could likewise obtain the scattering amplitudes
\begin{figure}[t]
\centering
\par
\begin{center}
\includegraphics[scale=0.35]{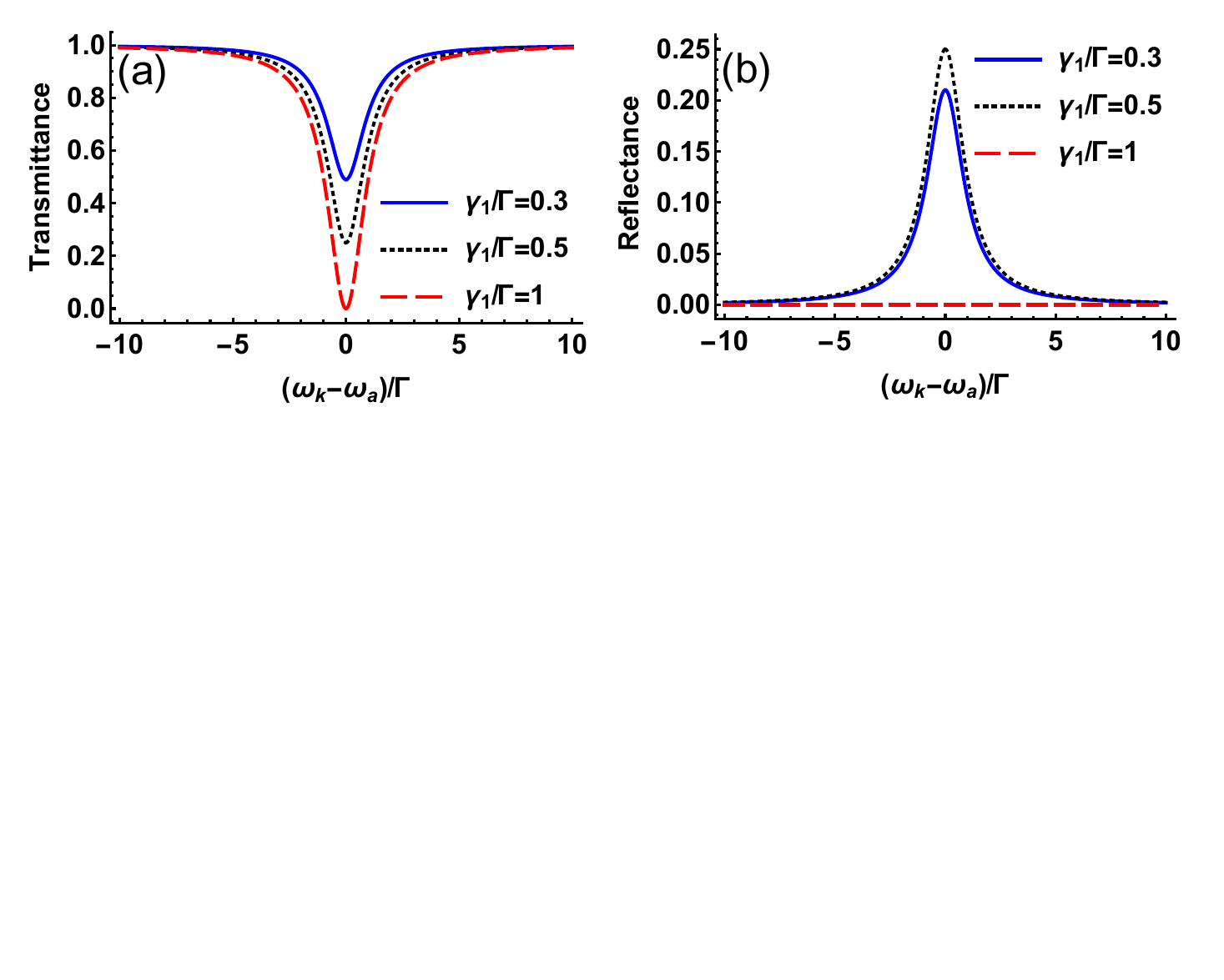}
\end{center}
\par
\renewcommand\figurename{\textbf{FIG.}}
\vspace{-40.0 mm}\caption{The transmittance $\left\vert \bar{t}_{k}\right\vert ^{2}$ (a) and
reflectance $\left\vert \bar{r}_{k}\right\vert ^{2}$ (b) plotted as functions
of ($\omega_{k}-\omega_{a}$)/$\Gamma$ for different values of $\gamma
_{1}/\Gamma$, with $\kappa/\Gamma=1$.}%
\label{Fig2}%
\end{figure}
\begin{align}
\tilde{t}_{k}  &  =\frac{\omega_{k}-\omega_{a}+i\frac{\kappa-\left(
\gamma_{2}-\gamma_{1}\right)  }{2}}{\omega_{k}-\omega_{a}+i\frac{\kappa
+\Gamma}{2}},\label{13a}\\
\tilde{r}_{k}  &  =\frac{-i\sqrt{\gamma_{1}\gamma_{2}}}{\omega_{k}-\omega
_{a}+i\frac{\kappa+\Gamma}{2}}. \label{13b}%
\end{align}
We show transmittance $\left\vert \bar{t}_{k}\right\vert ^{2}$ and reflectance
$\left\vert \bar{r}_{k}\right\vert ^{2}$ as functions of $\left(  \omega
_{k}-\omega_{a}\right)  /\Gamma$ for different values of $\gamma_{1}/\Gamma$,
with $\kappa/\Gamma=1$ in Fig. {\ref{Fig2}, where the single photon is injected
from the left. It can be seen the probability $P=\left\vert \bar{t}%
_{k}\right\vert ^{2}+\left\vert \bar{r}_{k}\right\vert ^{2}\neq1$ because the
waveguide is coupled to a dissipative cavity. Interestingly, $P=0$ when
$\gamma_{1}/\Gamma=1$, which means the single photon is completely dissipated
into the environment, so that the right-going single photon can not be transmitted.
When the incident photon is far detunned from the cavity, there is almost no
interaction between them, and the photon is nearly completely transmitted
through the waveguide.

Since we consider the chiral coupling, the transmmitance is different for the
left-going single-photon. We have depicted $\left\vert \bar{t}_{k}\right\vert
^{2}$and $\left\vert \tilde{t}_{k}\right\vert ^{2}$ as functions of
$\gamma_{1}/\Gamma$, with $\omega_{k}=\omega_{a}$ in Fig.\:\:\hyperref[Fig3]{3} for different
$\kappa/\Gamma$. For $\kappa=\Gamma$, the nonrecipoal effect is quite obvious
since the coupling between the waveguide and the cavity and its dissipation
has the same magnitude, as shown in Fig.\:\:\hyperref[Fig3]{3(a)}. On the other hand,
when the dissipation is negligible compared to $\Gamma$, there is almost no
nonrecipoal effect, as shown in Fig.\:\:\hyperref[Fig3]{3(b)}, coinciding with the result in \cite{36}. This is because the transmittances
are almost the same for $\bar{t}_{k}$ and $\tilde{t}_{k}$, according to Eqs.
(\ref{11a}) and (\ref{13a}). On the contrary, if the coupling $\Gamma$ is
negeligible compared to dissipation, the single photon injected from both
sides will be totally transmitted (depicted in Fig.\:\:\hyperref[Fig3]{3(c)}) because
the waveguide is weakly coupled to the cavity.

\begin{figure}[t]
\centering
\vspace{-0.0mm}\includegraphics[scale=0.35]{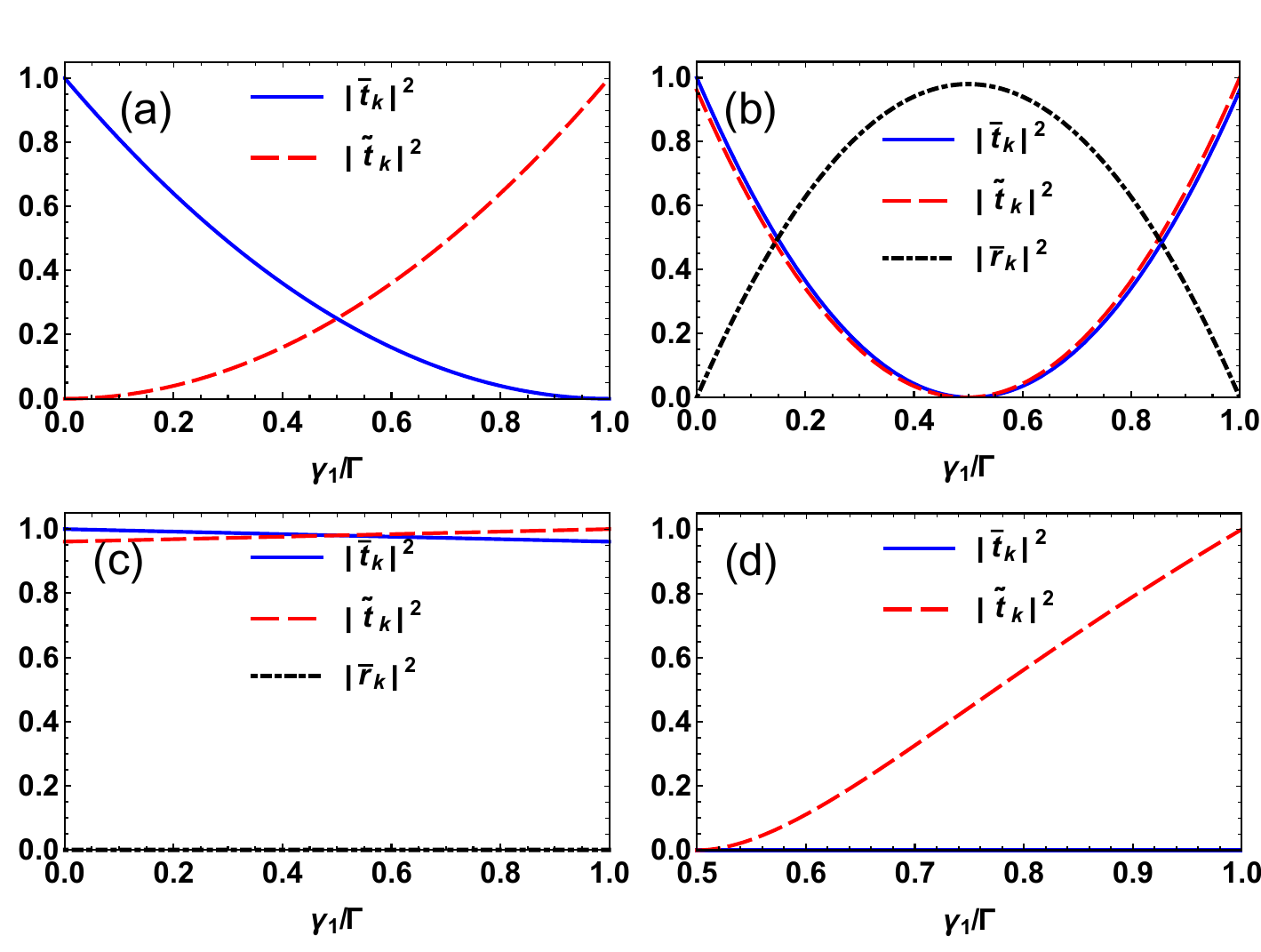}
\renewcommand\figurename{\textbf{FIG.}}
\vspace{-0.0mm}\caption{The transmittance and reflectance
plotted as functions of $\gamma_{1}/\Gamma$ for given $\kappa/\Gamma=1$ in
(a), $\kappa/\Gamma=0.01$ in (b), $\kappa/\Gamma=100$ in (c) and
$\kappa =\left( \gamma _{1} -\gamma _{2}
\right) \geqslant 0$ in (d), with $\omega_{k}=\omega_{a}$.}%
\label{Fig3}%
\end{figure}

As can be seen in Fig.\:\:\hyperref[Fig3]{3(a)}, an ideal photon diode is realized when
$\gamma_{1}/\Gamma=1$ or $0$, where the right-going (left-going) single photon
is completely transmitted while the left-going (right-going) one is totally
blocked. This is because $\left\vert \tilde{t}_{k}\right\vert ^{2}=1,$
$\left\vert \bar{t}_{k}\right\vert ^{2}=0$ when $\gamma_{1}/\Gamma=1$, and
$\left\vert \tilde{t}_{k}\right\vert ^{2}=0,$ $\left\vert \bar{t}%
_{k}\right\vert ^{2}=1$ when $\gamma_{1}/\Gamma=0$. Such chiral couplings can be realized as proposed in \cite{dimer}. Generally, $\left\vert
\bar{t}_{k}\right\vert ^{2}=0$ when $\kappa =\left( \gamma _{1} -\gamma _{2}
\right) \neq 0$, and
$\left\vert \tilde{t}_{k}\right\vert ^{2}=0$ when $\kappa =\left( \gamma _{2} -\gamma _{1}
\right) \neq 0$, which means diode can only be realized when dissipation presents for chiral coupling $\gamma_1\neq\gamma_2$ here. We depicted the former case in Fig.\:\:\hyperref[Fig3]{3(d)}, where
a photon diode is clearly realized, $\left\vert \tilde{t}%
_{k}\right\vert ^{2}$ reaches unity when $\gamma_{1}/\Gamma=1$, so we take it as
an ideal photon diode.
We set $\kappa/\Gamma=0.01$ in Fig.\:\:\hyperref[Fig3]{3(b)}. Then the cavity can be
considered as a perfect cavity, where the single photons incident from the
left and right sides only have a phase difference in the transmission
direction. The transmission probability reaches zero when $\gamma_{1}%
=\gamma_{2}$. So that total reflection occurs, blocking the transmission of
single photons.

\section{Two-Photon Scattering And Its Diode Effect}\label{IV}

We now consider the scattering dynamics in the double-excitation subspace,
searching for its diode effect. When two photons are injected from the left
side of the waveguide, the incident state reads
\begin{equation}
\left\vert \psi_{in}\right\rangle =\int\int dx_{1}dx_{2}\varphi_{k}\left(
x_{1,}x_{2}\right)  \frac{1}{\sqrt{2}}\hat{r}^{\dag}\left(  x_{1}\right)
\hat{r}^{\dag}\left(  x_{2}\right)  \left\vert \oslash\right\rangle
,\label{14}%
\end{equation}
where $\varphi_{k}\left(  x_{1,}x_{2}\right)  =\left(  e^{ik_{1}x_{1}+ik_{2}x_{2}%
}+e^{ik_{1}x_{2}+ik_{2}x_{1}}\right)  /\left(  2\sqrt{2}\pi\right)  $. The
scattering state can be obtained by solving the stationary Schr\"{o}dinger
equation (see the Appendix). The result reads \cite{40,41,42}
\begin{align}
\left\vert \psi\right\rangle  &  =\int\int dx_{1}dx_{2}\psi_{tt}\left(
x_{1},x_{2}\right)  \frac{1}{\sqrt{2}}\hat{r}^{\dag}\left(  x_{1}\right)
\hat{r}^{\dag}\left(  x_{2}\right)  \left\vert \oslash\right\rangle
\nonumber\\
&  +\int\int dx_{1}dx_{2}\psi_{rr}\left(  x_{1},x_{2}\right)  \frac{1}%
{\sqrt{2}}\hat{l}^{\dag}\left(  x_{1}\right)  \hat{l}^{\dag}\left(
x_{2}\right)  \left\vert \oslash\right\rangle \nonumber\\
&  +\int\int dx_{1}dx_{2}\psi_{rt}\left(  x_{1},x_{2}\right)  \hat{l}^{\dag
}\left(  x_{2}\right)  \hat{r}^{\dag}\left(  x_{1}\right)  \left\vert
\oslash\right\rangle ,
\end{align}
where
\begin{align}
\psi_{tt}\left(  x_{1},x_{2}\right)   &  =\varphi_{k}\left(  x_{1}%
,x_{2}\right)  \bar{t}_{k_{1}}\bar{t}_{k_{2}}\nonumber\\
&  +\frac{\gamma_{1}^{2}}{\Gamma^{2}}De^{i\omega x_{c}}e^{\left[  i\left(
\omega-2\omega_{a}\right)  -\left(  \kappa+\Gamma\right)  \right]
\frac{\left\vert x\right\vert }{2}},\label{16}\\
\psi_{rr}\left(  x_{1},x_{2}\right)   &  =\varphi_{k}\left(  -x_{1}%
,-x_{2}\right)  \bar{r}_{k_{1}}\bar{r}_{k_{2}}\nonumber\\
&  +\frac{\gamma_{1}\gamma_{2}}{\Gamma^{2}}De^{-i\omega x_{c}}e^{\left[
i\left(  \omega-2\omega_{a}\right)  -\left(  \kappa+\Gamma\right)  \right]
\frac{\left\vert x\right\vert }{2}},\\
\psi_{rt}\left(  x_{1},x_{2}\right)   &  =\frac{1}{4\pi}\left[  \varphi
_{k}\left(  x_{1},-x_{2}\right)  \bar{t}_{k_{1}}\bar{r}_{k_{2}}+\varphi
_{k}\left(  -x_{2},x_{1}\right)  \bar{r}_{k_{1}}\bar{t}_{k_{2}}\right]
\nonumber\\
&  +\frac{\sqrt{2\gamma_{1}^{3}\gamma_{2}}}{\Gamma^{2}}De^{-i\frac{\omega}%
{2}x}e^{\left[  i\left(  \omega-2\omega_{a}\right)  -\left(  \kappa
+\Gamma\right)  \right]  \left\vert x_{c}\right\vert },\\
\bar{t}_{k_{i}} &  =\frac{\omega_{k_{i}}-\omega_{a}+i\frac{\kappa-\left(
\gamma_{1}-\gamma_{2}\right)  }{2}}{\omega_{k_{i}}-\omega_{a}+i\frac
{\kappa+\Gamma}{2}},\\
\bar{r}_{k_{i}} &  =\tilde{r}_{k_{i}}=\frac{-i\sqrt{\gamma_{1}\gamma_{2}}%
}{\omega_{k_{i}}-\omega_{a}+i\frac{\kappa+\Gamma}{2}}.
\end{align}

$D$ is a function of $\Gamma$, $U$, $\kappa$, $\omega_{k_{i}}$ and $\omega_{a}$ (see the Appendix). $\psi_{tt}\left(  x_{1},x_{2}\right)  $ represents the
wavefunction for two transmitted photons. $\psi_{rr}\left(  x_{1}%
,x_{2}\right)  $ is the wavefunction of the two reflected photons. $\psi
_{rt}\left(  x_{1},x_{2}\right)  $ is the wavefunction for one reflected
photon and one transmitted photon. The center-of-mass coordinate
$x_{c}=\left(  x_{1}+x_{2}\right)  /2$ and the relative coordinate
$x=x_{2}-x_{1}$. $\omega=\omega_{k_{1}}+\omega_{k_{2}}$. $\bar{t}_{k_{i}}$
(for $i=1,2$) is the transmitted amplitude and $\bar{r}_{k_{i}}$ is the reflected
amplitude for the single-photon case. Now the scattering amplitudes consist of
a plane wave part and a bound state part which decays with $x_{c}$ or $x$.
Such bound states between two photons are induced by their interaction with
the nonlinear cavity, which will produce interesting physical effects as
discussed later.

When two photons enter the cavity from the right side of the waveguide, the
scattering state can also be obtained in a similar way as
\begin{align}
\tilde{\psi}_{tt}\left(  x_{1},x_{2}\right)   &  =\varphi_{k}\left(
-x_{1},-x_{2}\right)  \tilde{t}_{k_{1}}\tilde{t}_{k_{2}}\nonumber\\
&  +\frac{\gamma_{2}^{2}}{\Gamma^{2}}De^{-i\omega x_{c}}e^{\left[  i\left(
\omega-2\omega_{a}\right)  -\left(  \kappa+\Gamma\right)  \right]
\frac{\left\vert x\right\vert }{2}},\label{23}\\
\tilde{\psi}_{rr}\left(  x_{1},x_{2}\right)   &  =\varphi_{k}\left(
x_{1},x_{2}\right)  \tilde{r}_{k_{1}}\tilde{r}_{k_{2}}\nonumber\\
&  +\frac{\gamma_{1}\gamma_{2}}{\Gamma^{2}}De^{i\omega x_{c}}e^{\left[
i\left(  \omega-2\omega_{a}\right)  -\left(  \kappa+\Gamma\right)  \right]
\frac{\left\vert x\right\vert }{2}},\\
\tilde{\psi}_{rt}\left(  x_{1},x_{2}\right)   &  =\frac{1}{4\pi}\left[
\varphi_{k}\left(  -x_{1},x_{2}\right)  \tilde{t}_{k_{1}}\tilde{r}_{k_{2}%
}+\varphi_{k}\left(  x_{2},-x_{1}\right)  \tilde{r}_{k_{1}}\tilde{t}_{k_{2}%
}\right] \nonumber\\
&  +\frac{\sqrt{2\gamma_{1}\gamma_{2}^{3}}}{\Gamma^{2}}De^{-i\frac{\omega}%
{2}x}e^{\left[  i\left(  \omega-2\omega_{a}\right)  -\left(  \kappa
+\Gamma\right)  \right]  \left\vert x_{c}\right\vert },\\
\tilde{t}_{k_{i}}  &  =\frac{\omega_{k_{i}}-\omega_{a}+i\frac{\kappa+\left(
\gamma_{1}-\gamma_{2}\right)  }{2}}{\omega_{k_{i}}-\omega_{a}+i\frac
{\kappa+\Gamma}{2}}.
\end{align}

\begin{figure}[t]
\centering
\hspace{-4mm} \includegraphics[scale=0.39]{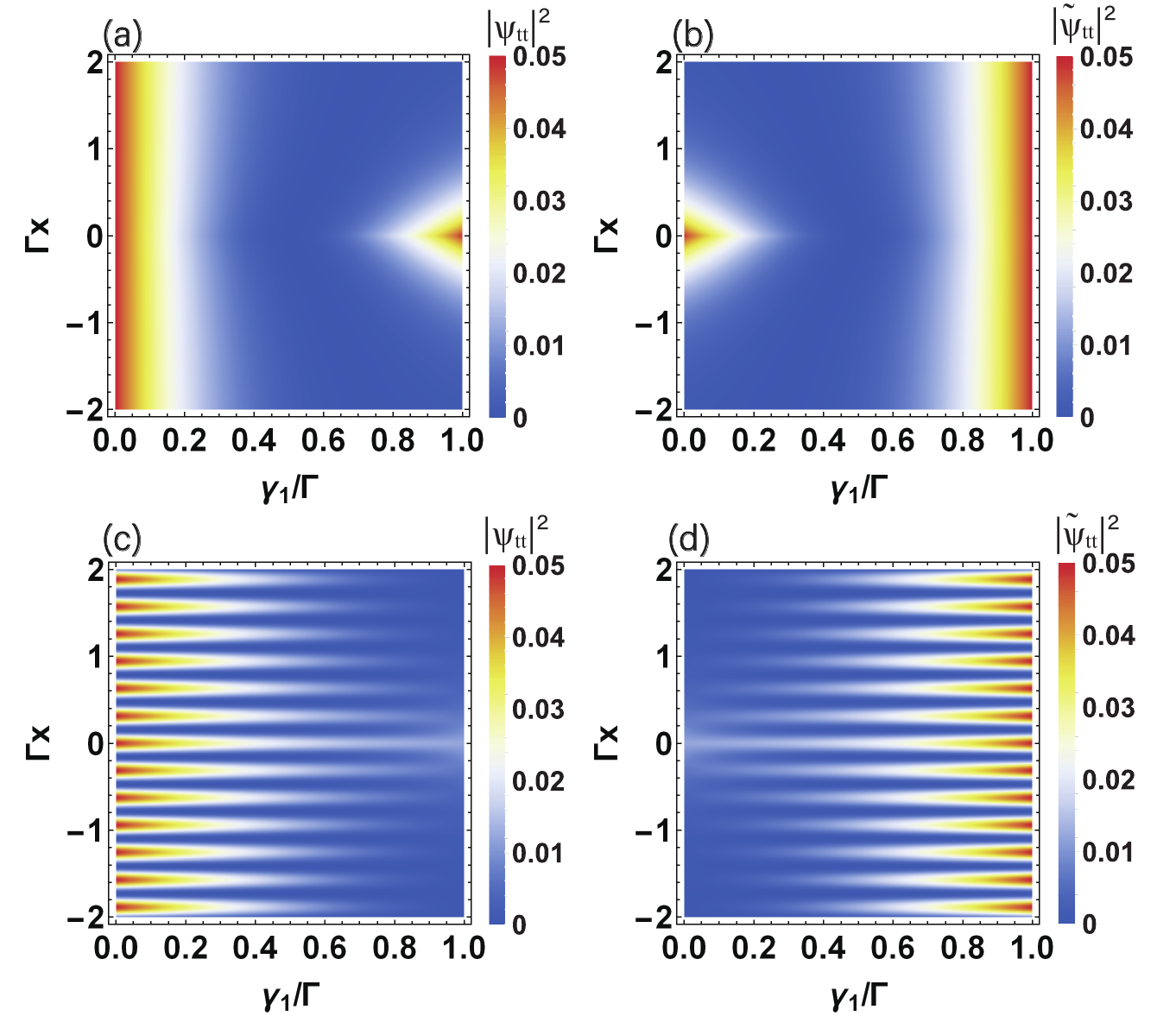}
\renewcommand\figurename{\textbf{FIG.}}
\caption{(a) and (b) are plots of $\left\vert \psi_{tt}\left(  x_{1},x_{2}\right)
\right\vert ^{2}$ and $\left\vert \tilde{\psi}_{tt}\left(  x_{1},x_{2}\right)
\right\vert ^{2}$, respectively, when $\omega_{k_{1}}=\omega_{k_{2}}%
=\omega_{a}$. (c) and (d) are plots of $\left\vert \psi_{tt}\left(
x_{1},x_{2}\right)  \right\vert ^{2}$ and $\left\vert \tilde{\psi}_{tt}\left(
x_{1},x_{2}\right)  \right\vert ^{2}$, respectively, when $\omega_{k_{1}%
}=\omega_{a}$, $\omega_{k_{2}}=\omega_{a}+2U$. Other parameters $U/\Gamma=10$
and $\kappa/\Gamma=1$. All parameters are in units of $\Gamma$.}%
\label{Fig4}%
\end{figure}

\subsection{Two-photon scattering for $\kappa$ approaching $\Gamma
$}\label{A}

We focus on the case where two photons are transmitted by analyzing
$\left\vert \psi_{tt}\left(  x_{1},x_{2}\right)  \right\vert ^{2}$. The ratio
of cavity dissipation to the coupling strength will significantly affect the
scattering results. We first consider the case where they are of the same
order. We chose the resonance condition $\omega_{k_{i}}=\omega_{a}$ and depict
$\left\vert \psi_{tt}\left(  x_{1},x_{2}\right)  \right\vert ^{2}$ from the
right-going photons and $\left\vert \tilde{\psi}_{tt}\left(  x_{1}%
,x_{2}\right)  \right\vert ^{2}$ from the left-going photons, in Figs.\:\:\hyperref[Fig4]{4(a)} and\:\:\hyperref[Fig4]{4(b)} respectively. When $\gamma_{1}/\Gamma=0$,
the right-going photons do not interact with the cavity and are totally
transmitted in the form of plane waves. Meanwhile, the left-going photons
interacting with the cavity are almost fully dissipated since $\kappa=\Gamma$,
as can be seen in Figs.\:\:\hyperref[Fig4]{4(a)} and\:\:\hyperref[Fig4]{4(b)}. However, unlike
the single-photon case, the transmittance does not vanish due to the bound
state in Eq. (\ref{23}). It decays with the distance between two photons as%

\begin{equation}
\frac{32U^{2}}{\pi^{2}\left(  \kappa+\Gamma\right)  ^{4}\left(  4U^{2}+\left(
\kappa+\Gamma\right)  ^{2}\right)  }e^{-\left\vert x\right\vert \left(
\kappa+\Gamma\right)  }, \label{30}%
\end{equation}
indicating a bunching effect. There is a peak located at $\Gamma x=0$ for the
transmittance. Now the photon diode effect is realized at $x\gg1/\left( \kappa +\Gamma \right)$. As
$\gamma_{1}$ increases, the right-going photons begin to interact with the
cavity, so that they can be dissipated into the environment, transmitted or
reflected. When $\gamma_{1}/\Gamma=1$, the left-going photons will not
interact with the cavity, and symmetric dynamical process can be observed as
$\gamma_{1}/\Gamma=0$ since $\gamma_{1}$ and $\gamma_{2}$ are swapped.

Then we consider the two-photon resonance condition $\omega_{k_{1}}=\omega_{a}$, $\omega
_{k_{2}}=\omega_{a}+2U$. When $\gamma_{1}/\Gamma=0$, the
right-going photons do not interact with the cavity and are completely
transmitted in the form of a plane wave. The interference pattern in Fig.\:\:\hyperref[Fig4]{4(c)} are formed
just by the two-photon plane wave $\varphi_{k}\left(  x_{1},x_{2}\right)$ as
$P\left(  x\right)  =\left\vert \varphi_{k}\left(  x_{1,}x_{2}\right)
\right\vert ^{2}=\cos^{2}\left(  Ux\right)  /\left(  2\pi^{2}\right)  $. As
$\gamma_{1}$ increases, a two-photon bound state is formed mediated by the
cavity. It contributes to the transmittance $\left\vert \psi_{tt}\left(
x_{1},x_{2}\right)  \right\vert ^{2}$. When $\gamma_{1}/\Gamma=1$, the plane
wave part has no contribution to $\left\vert \psi_{tt}\left(  x_{1}%
,x_{2}\right)  \right\vert ^{2}$, the same as the single-photon resonance case. However,
the contribution from the remaining bound state part%

\begin{equation}
\frac{32U^{2}}{\pi^{2}\left(  \kappa+\Gamma\right)  ^{4}\left(  16U^{2}%
+\left(  \kappa+\Gamma\right)  ^{2}\right)  }e^{-\left\vert x\right\vert
\left(  \kappa+\Gamma\right)  } \label{31}%
\end{equation}
enables the photons to be transmitted. Its probability decays exponentially
with their relative coordinates $x$. The decay width is $\kappa+\Gamma$
($\upsilon_{c}=1$). The dynamics of the left-going photons are the same as the
right-going photons if we interchange $\gamma_{1}$ and $\gamma_{2}$, as shown
in Fig.\:\:\hyperref[Fig4]{4(d)}.

\begin{figure}[t]
\centering
\vspace{0.0mm}\includegraphics[scale=0.35]{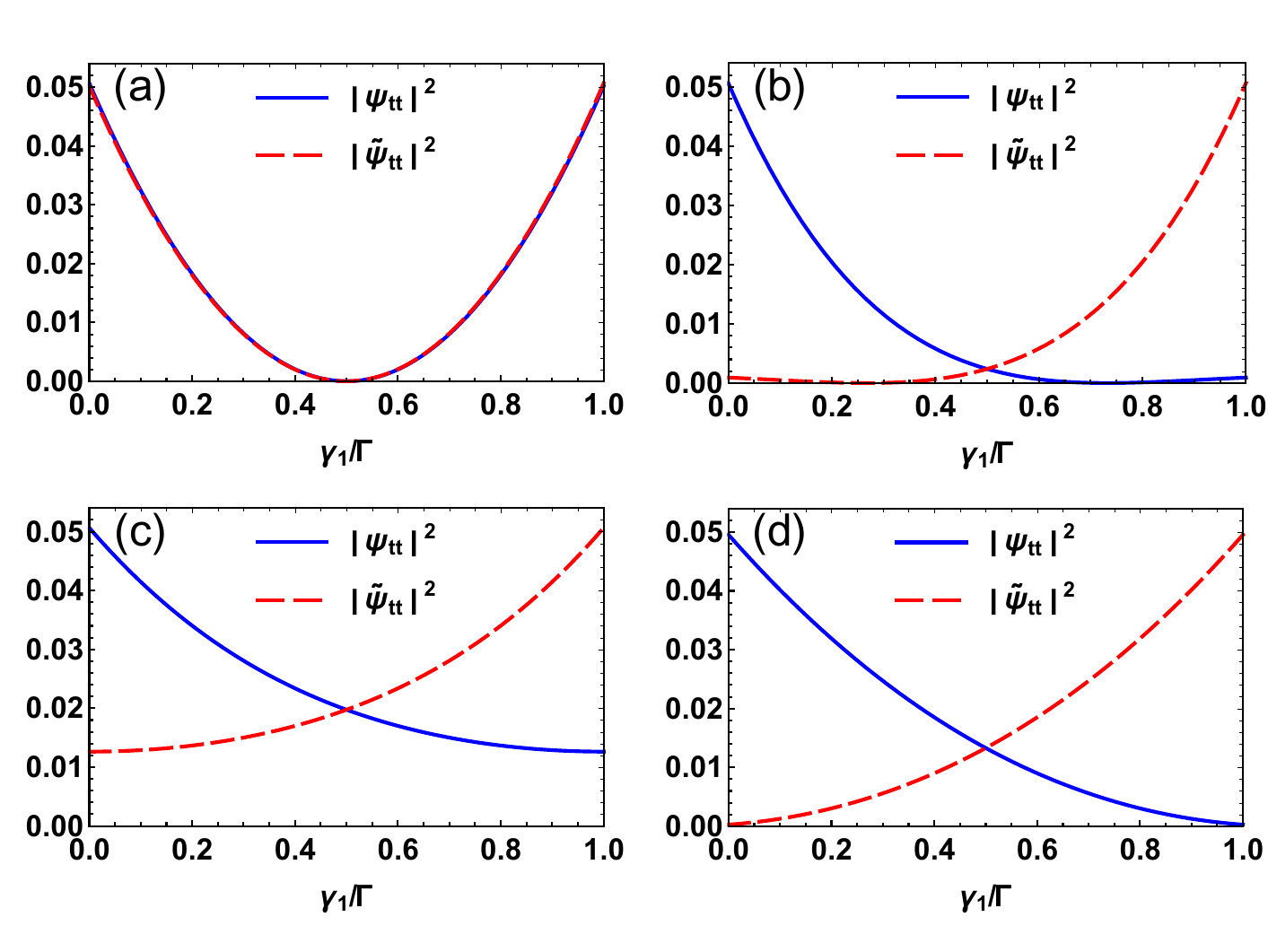}
\renewcommand\figurename{\textbf{FIG.}}
\caption{$\left\vert \psi
_{tt}\left(  x_{1},x_{2}\right)  \right\vert ^{2}$ and $\left\vert \tilde
{\psi}_{tt}\left(  x_{1},x_{2}\right)  \right\vert ^{2}$ plotted as a function
of $\gamma_{1}/\Gamma$ with $U/\Gamma=10$, $\kappa/\Gamma=1$. The upper panel
deals with the single-photon resonance case $\omega_{k_{1}}=\omega_{k_{2}%
}=\omega_{a}$. (a) $\Gamma x=0$ and (b) $\Gamma x=2$. The lower panel deals
with the two-photon resonance case $\omega_{k_{1}}=\omega_{a}$, $\omega
_{k_{2}}=\omega_{a}+2U$. (c) $\Gamma x=0$ and (d) $\Gamma x=1.9$.}%
\label{Fig5}%
\end{figure}\ \ \

\subsection{Working area of the diode}\label{B}

As can be seen from Eqs. (\ref{16}) and (\ref{23}), $\psi_{tt}\left(
x_{1},x_{2}\right)  $ and $\tilde{\psi}_{tt}\left(  x_{1},x_{2}\right)  $
vanish at specific $\Gamma x$ for specific parameters $\gamma_{1}/\Gamma$,
$\kappa/\Gamma$ and $U/\Gamma$. In order to find the working area of the
diode, we plotted them under the single-photon resonance case in Figs.\:\:\hyperref[Fig5]{5(a)} and\:\:\hyperref[Fig5]{5(b)}. We first consider the bound state area $\Gamma x=0$ at
$\gamma_{1}/\Gamma=1$ with $\Gamma=\kappa$. The left-going photons do not
interact with the cavity, so that they are transmitted. The right-going
photons are coupled to the cavity, and only the bound state part contributes
to the transmittance. It is maximized at $\Gamma x=0$, which are almost the
same as the left-going photons at this specific position, as shown in Fig.\:\:\hyperref[Fig5]{5(a)}, where the diode effects almost vanishes. However, it recovers
at the region $\left(  \kappa+\Gamma\right)  x\gg1$. This can be seen form
Fig.\:\:\hyperref[Fig5]{5(b)}, where $\Gamma x=2$. Interestingly, the minimum of
$\left\vert \psi_{tt}\left(  x_{1},x_{2}\right)  \right\vert ^{2}$ does not
present at $\gamma_{1}/\Gamma=1$. This means the coherent superposition of the
plane-wave part and the bound state part can further reduce the transmittance.
To have a rigorous diode effect, $\left\vert \psi_{tt}\left(  x_{1}%
,x_{2}\right)  \right\vert ^{2}$ should vanish. This can be done if both parts
cancel each other out. As can be seen in Eq. (\ref{16}), $\left\vert \psi
_{tt}\left(  x_{1},x_{2}\right)  \right\vert ^{2}$ vanishes if $U\rightarrow
\infty$ at%

\begin{equation}
\left\vert x\right\vert =\frac{2}{\kappa+\Gamma}\ln\left[  \frac{4\gamma
_{1}^{2}}{\left(  \Gamma+\kappa-2\gamma_{1}\right)  ^{2}}\right]  . \label{33}%
\end{equation}
This equation is satisfied only for $\left(  \kappa+\Gamma\right)
/4\leqslant\gamma_{1}\leqslant\Gamma$. The transmitted two-photon incidents from
the left vanishes at such points. We plotted them for $\kappa=\Gamma$ in Fig.\:\:\hyperref[Fig9]{6(a)}. We also solved $\left\vert \psi
_{tt}\left(  x_{1},x_{2}\right)  \right\vert ^{2}\approx 0$ numerically at $U/\Gamma=10$, and find the results coincide with Eq. (\ref{33}). $\Gamma x\rightarrow\infty$ at $\gamma_{1}/\Gamma=1$,
as shown in Fig.\:\:\hyperref[Fig9]{6(a)}. Interestingly, $\Gamma x$
decreases as $\gamma_{1}/\Gamma$ decreases, and reaches $0$ at $\gamma
_{1}/\Gamma=1/2$. This means the maximum point of the bound state prevents the
two photons to be transmitted under reciprocal coupling $\gamma_{1}%
/\Gamma=\gamma_{2}/\Gamma$.

\begin{figure}[t]
\centering
\includegraphics[scale=0.35]{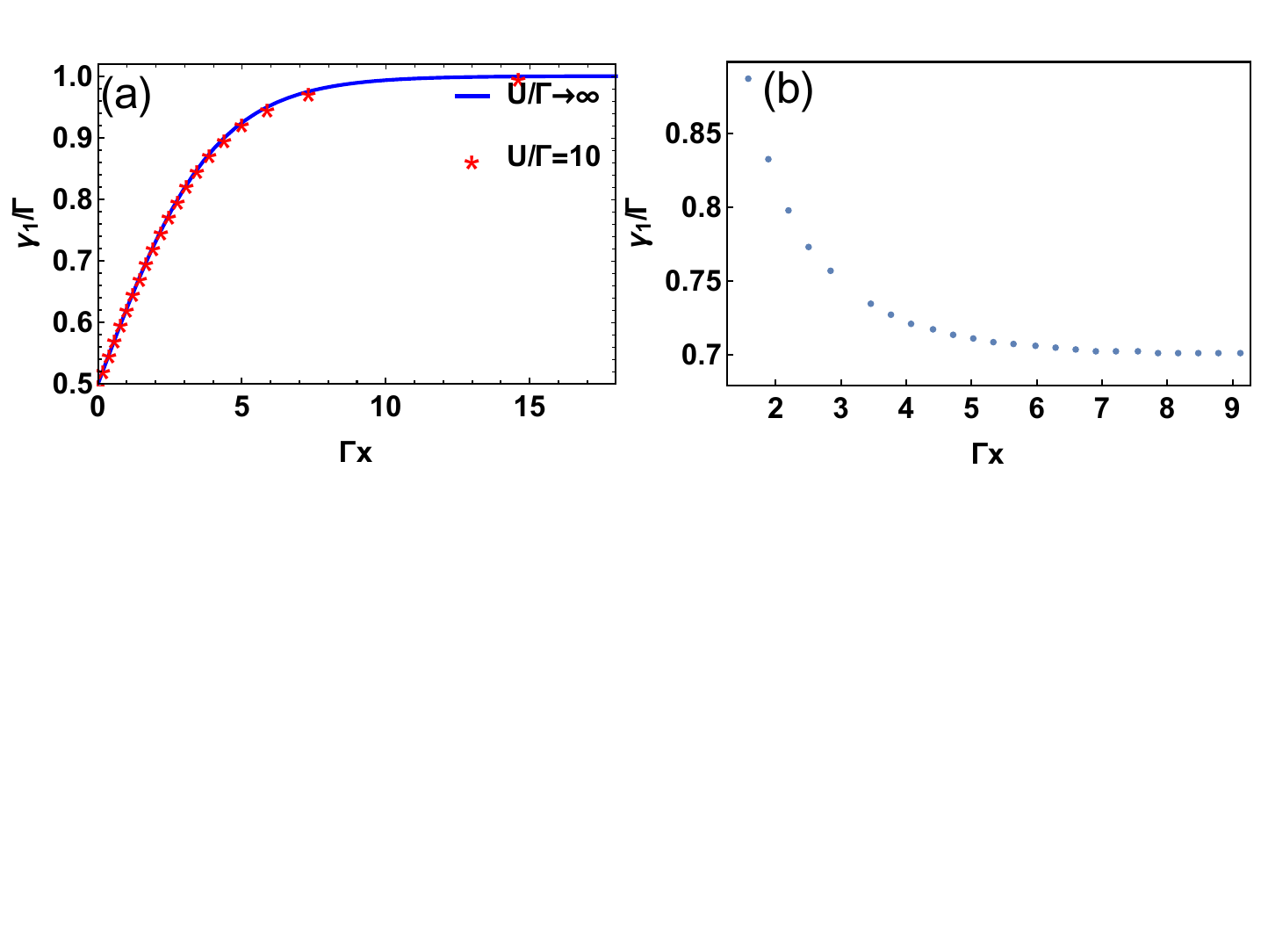}
\renewcommand\figurename{\textbf{FIG.}}
\vspace{-30mm}\caption{Working area of the diode satisfying
$\left\vert \psi_{tt}\left(  x_{1},x_{2}\right)  \right\vert ^{2}=0$. (a) The
single-photon resonance case $\omega_{k_{1}}=\omega_{k_{2}}=\omega_{a}$.
$\kappa/\Gamma=1$. (b) The two-photon
resonance case $\omega_{k_{1}}=\omega_{a}$, $\omega_{k_{2}}=\omega_{a}+2U$.
$U/\Gamma=10$ and $\kappa/\Gamma=0.4$.}%
\label{Fig9}%
\end{figure}

Next we consider the two-photon resonance case. The non-reciprocity effect is
obvious even at $\Gamma x=0$, as shown in Fig.\:\:\hyperref[Fig5]{5(c)}. The diode effect
is recovered at $\left(  \kappa+\Gamma\right)  x\gg1$, as depicted in Fig.\:\:\hyperref[Fig5]{5(d)}. If $\gamma_{1}$ and $\gamma_{2}$ are interchanged, then the
dynamics for the right-going photons and left-going photons are also
interchanged. To find the working area of the two-photon diode, we solve
$\left\vert \psi_{tt}\left(  x_{1},x_{2}\right)  \right\vert ^{2}=0$ to get%
\begin{align}
\left\vert x\right\vert  &  =\frac{2}{\Gamma+\kappa}\ln\left[  \frac
{2\gamma_{1}^{2}}{\left(  2\gamma_{1}-\kappa-\Gamma\right)  \left(
\Gamma+\kappa\right)  }\right]  ,\label{34}\\
\tan\left[  U\left\vert x\right\vert \right]   &  =\frac{\Gamma+\kappa
-2\gamma_{1}}{4U}. \label{35}%
\end{align}
Now the parameters $\gamma_{1}$, $\kappa$, $\Gamma$, $U$ has to satisfy Eqs.
(\ref{34}) and (\ref{35}). Meanwhile, $0.5<\gamma_{1}/\Gamma\leqslant1$ to
satisfy Eq. (\ref{34}). Now $\left\vert \psi_{tt}\left(  x_{1},x_{2}\right)
\right\vert ^{2}$ vanishes at specific $\Gamma x$ and $\gamma_{1}/\Gamma$ when
$U/\Gamma$ and $\kappa/\Gamma$ are fixed, as shown in Fig.\:\:\hyperref[Fig9]{6(b)}.
Interestingly, $\Gamma x\rightarrow\infty$ when $\gamma_{1}=\left(
\kappa+\Gamma\right)  /2$.

\begin{figure}[ptbh]
\centering
\includegraphics[scale=0.4]{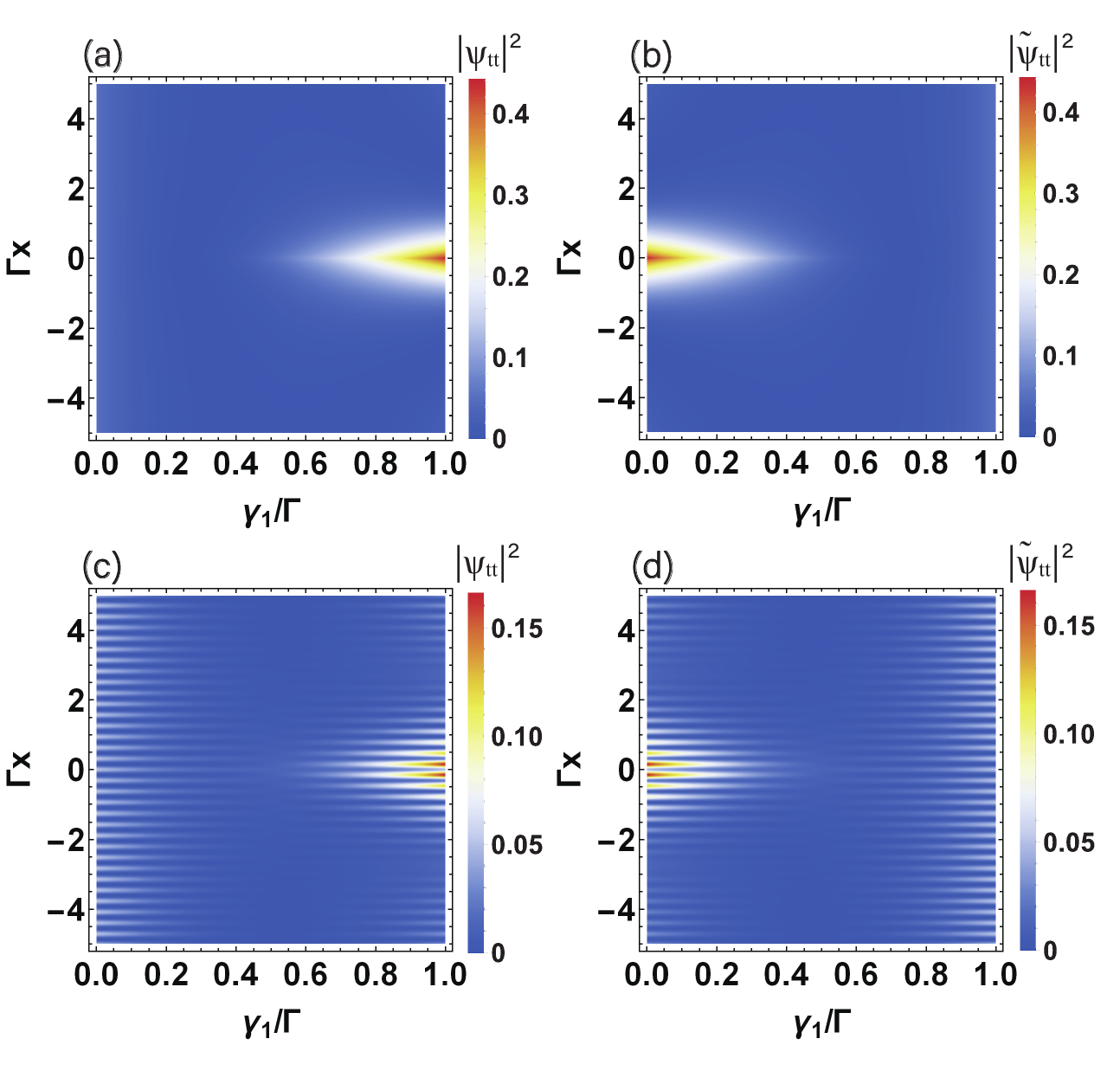}
\renewcommand\figurename{\textbf{FIG.}}
\caption{(a) and (b) are plots of
$\left\vert \psi_{tt}\left(  x_{1},x_{2}\right)  \right\vert ^{2}$ and
$\left\vert \tilde{\psi}_{tt}\left(  x_{1},x_{2}\right)  \right\vert ^{2}$,
respectively, when $\omega_{k_{1}}=\omega_{k_{2}}=\omega_{a}$. (c) and (d) are
plots of $\left\vert \psi_{tt}\left(  x_{1},x_{2}\right)  \right\vert ^{2}$
and $\left\vert \tilde{\psi}_{tt}\left(  x_{1},x_{2}\right)  \right\vert ^{2}%
$, respectively, when $\omega_{k_{1}}=\omega_{a}$, $\omega_{k_{2}}=\omega
_{a}+2U$. Other parameters are set as $U/\Gamma=10$ and $\kappa/\Gamma=0.01$,
All parameters are in units of $\Gamma$.}%
\label{Fig6}%
\end{figure}

\subsection{Two-photon scattering for $\kappa\ll\Gamma$ and $\kappa \gg \Gamma $}\label{C}

To emphasis the effect of dissipation, we compare the case discussed in Secs. A and B with the cases $\kappa\ll\Gamma$ and $\kappa\gg\Gamma$. First we consider $\kappa\ll\Gamma$ under the single-photon resonance case $\omega_{k_{1}}=\omega_{k_{2}}=\omega_{a}$, and depicted $\left\vert \psi_{tt}\left(
x_{1},x_{2}\right)  \right\vert ^{2}$ in Fig.\:\:\hyperref[Fig6]{7(a)}and $\left\vert
\tilde{\psi}_{tt}\left(  x_{1},x_{2}\right)  \right\vert ^{2}$ in Fig.\:\:\hyperref[Fig6]{7(b)}. When $\gamma_{1}/\Gamma=0$, the right-going photons do not
interact with the cavity and are completely transmitted in the form of plane
waves with $\left\vert \varphi_{k}\left(  x_{1,}x_{2}\right)  \right\vert
^{2}=1/\left(  2\pi^{2}\right)  $. When $\gamma_{1}/\Gamma=1$, the bound state
component in $\psi_{tt}\left(  x_{1},x_{2}\right)  $ reaches its maximum.
Because the lack of dissipation, the transmittance is much enhanced. Together
with the plane wave part, $\left\vert \psi_{tt}\left(  x_{1},x_{2}\right)
\right\vert ^{2}$ at $x=0$ reaches 0.442976, as shown in Fig.\:\:\hyperref[Fig6]{7(a)}.
The dynamics of the left-going photons is the same as the right-going photons
if we replace $\gamma_{1}$ with $\gamma_{2}$, which can be seen in Fig.\:\:\hyperref[Fig6]{7(b)}. Next we consider the two-photon resonance condition
$\omega_{k_{1}}=\omega_{a}$, $\omega
_{k_{2}}=\omega_{a}+2U$. Due to the lack of
dissipation, the amplitudes of the bound state part in $\left\vert \psi
_{tt}\left(  x_{1},x_{2}\right)  \right\vert ^{2}$ and $\left\vert \tilde
{\psi}_{tt}\left(  x_{1},x_{2}\right)  \right\vert ^{2}$ are enlarged,
as can be seen from Figs.\:\:\hyperref[Fig6]{7(c)} and\:\:\hyperref[Fig6]{7(d)}, by comparing with Figs.\:\:\hyperref[Fig4]{4(c)} and\:\:\hyperref[Fig4]{4(d)}.

\begin{figure}[t]
\centering
\vspace{-0.0 mm} \includegraphics[scale=0.35]{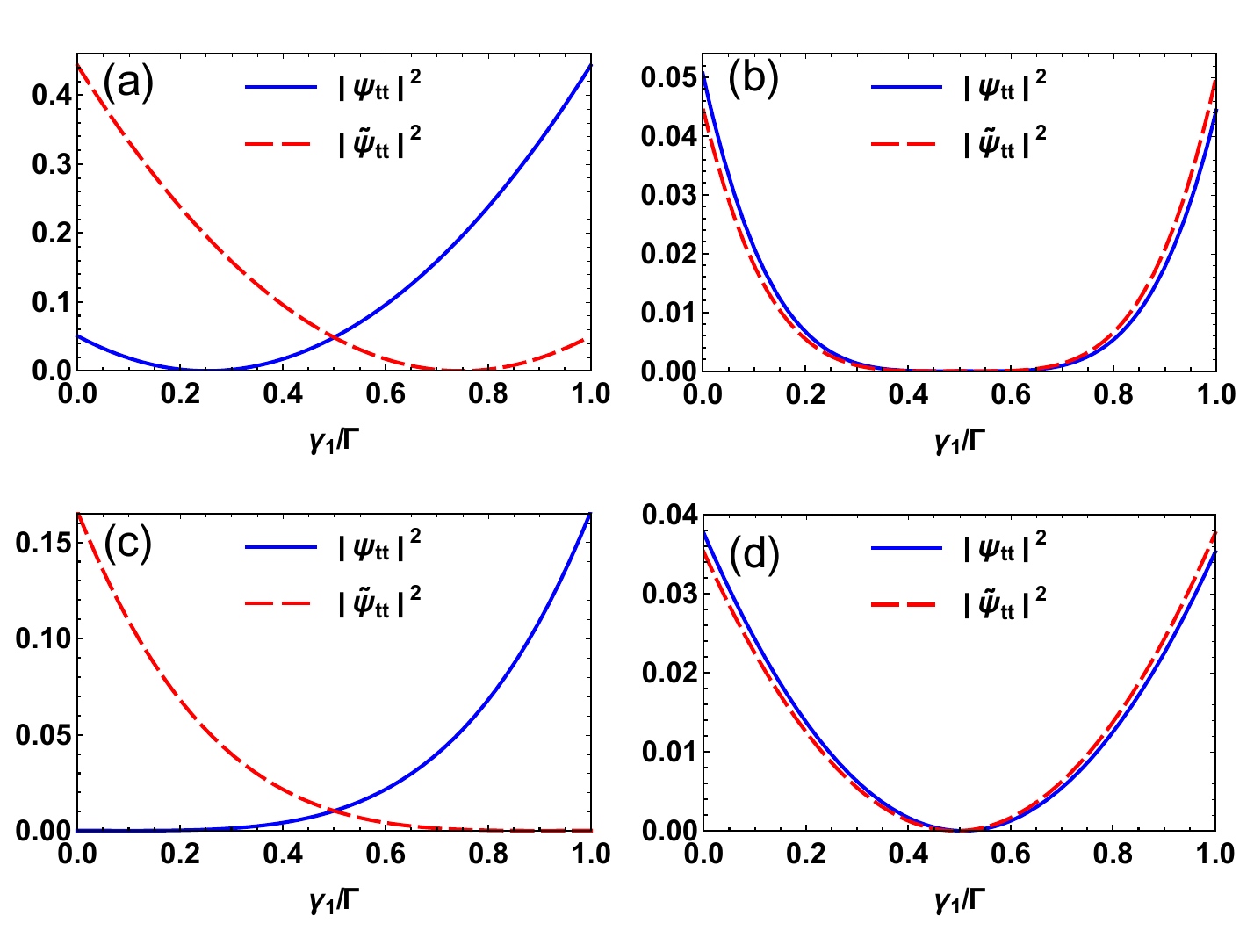}
\renewcommand\figurename{\textbf{FIG.}}
\caption{$\left\vert
\psi_{tt}\left(  x_{1},x_{2}\right)  \right\vert ^{2}$ and $\left\vert
\tilde{\psi}_{tt}\left(  x_{1},x_{2}\right)  \right\vert ^{2}$ plotted as a
function of $\gamma_{1}/\Gamma$, when $\omega_{k_{1}}=\omega_{k_{2}}%
=\omega_{a}$. (a) $\Gamma x=0$ and (b) $\Gamma x=5$. $\left\vert \psi
_{tt}\left(  x_{1},x_{2}\right)  \right\vert ^{2}$ and $\left\vert \tilde
{\psi}_{tt}\left(  x_{1},x_{2}\right)  \right\vert ^{2}$ plotted as a function
of $\gamma_{1}/\Gamma$, when $\omega_{k_{1}}=\omega_{a}$, $\omega_{k_{2}%
}=\omega_{a}+2U$. (c) $\Gamma x=0.15$ and (d) $\Gamma x=10$. Here,
$U/\Gamma=10$, $\kappa/\Gamma=0.01$.}%
\label{Fig7}%
\end{figure}

In order to find out the optimal region for the diode effect, we plotted
$\left\vert \psi_{tt}\left(  x_{1},x_{2}\right)  \right\vert ^{2}$ and
$\left\vert \tilde{\psi}_{tt}\left(  x_{1},x_{2}\right)  \right\vert ^{2}$ for
single-photon resonance case in Figs.\:\:\hyperref[Fig7]{8(a-b)} and for two-photon
resonance case in Figs.\:\:\hyperref[Fig7]{8(c-d)}. Since the dissipation is
negligible, the nonrecipol effect is only induced by the bound state part,
as can be found from Eqs. (\ref{16}) and (\ref{23}). First we consider the
peak of the bound state at $\Gamma x=0$ for the single-photon resonance case.
As can be seen in Fig.\:\:\hyperref[Fig7]{8(a)}, $\left\vert \psi_{tt}\left(
x_{1},x_{2}\right)  \right\vert ^{2}$ and $\left\vert \tilde{\psi}_{tt}\left(
x_{1},x_{2}\right)  \right\vert ^{2}$ are quite different. Although the bound state part is proportional to $\gamma_1$ in $\left\vert \psi_{tt}\left(  x_{1},x_{2}\right)  \right\vert ^{2}$, the
minimum of $\left\vert \psi_{tt}\left(  x_{1},x_{2}\right)  \right\vert ^{2}$
does not present at $\gamma_{1}/\Gamma=0$, because the coherent
superposition of the plane-wave part and bound state part reduces the
transmittance further. A diode effect can be realized at $\left\vert \Gamma
x\right\vert =0$ for $\gamma_{1}/\Gamma=1/4$ and $U\rightarrow\infty$,
according to Eq. (\ref{33}). This coincides with the results in Fig.\:\:\hyperref[Fig7]{8(a)} where $U/\Gamma=10$. The position for the diode effect $\left\vert \Gamma x\right\vert $
increases with $\gamma_{1}/\Gamma$, and finally reaches $\left(
2/\Gamma\right)  \ln4$ at $\gamma_{1}/\Gamma=1$. When $\Gamma x\gg1$, the
transmittance of the right-going and left-going photons are almost the same,
as can be seen in Fig.\:\:\hyperref[Fig7]{8(b)}.

Next we consider the two-photon resonance case, the nonreciprocal effect is particularly pronounced in the bound state part. A diode effect is realized when $\Gamma
x\ll1$, as can be seen in Fig.\:\:\hyperref[Fig7]{8(c)}. However, it vanishes when
$\Gamma x\gg1$, as shown in Fig.\:\:\hyperref[Fig7]{8(d)}. The diode effect becomes significantly more pronounced at the two-photon level \cite{36}.

\begin{figure}[t]
\centering
\includegraphics[scale=0.35]{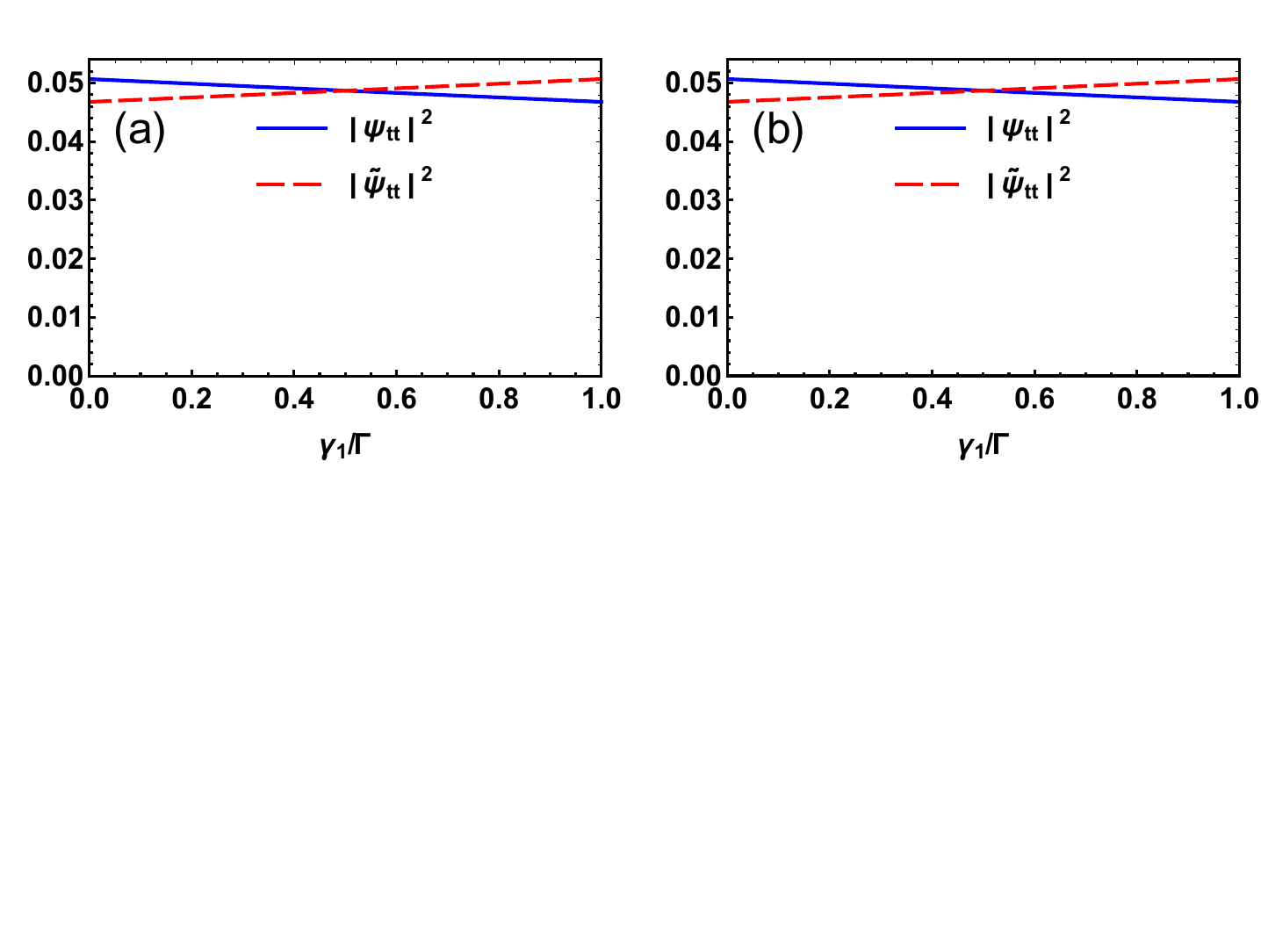}
\renewcommand\figurename{\textbf{FIG.}}
\vspace{-30.0 mm}\caption{$\left\vert \psi_{tt}\left(
x_{1},x_{2}\right)  \right\vert ^{2}$ and $\left\vert \tilde{\psi}_{tt}\left(
x_{1},x_{2}\right)  \right\vert ^{2}$ plotted as a function of $\gamma
_{1}/\Gamma$, when $\omega_{k_{1}}=\omega_{k_{2}}=\omega_{a}$, (a) $\Gamma
x=0$ and (b) $\Gamma x=5$. Here, $U /\Gamma=10$, $\kappa/\Gamma=100$. }%
\label{Fig8}%
\end{figure}

When $\Gamma\ll \kappa$, the nonlinear cavity can be regarded as an environment, weakly
coupled to the waveguide, so that the photon are almost propagated freely. It
can be seen from Fig.\:\:\hyperref[Fig8]{9(a)} that $\left\vert \psi_{tt}\left(
x_{1},x_{2}\right)  \right\vert ^{2}$ decreases as $\gamma_{1}/\Gamma$
increases, due to the photon loss by coupling to the dissipative cavity. It
can be seen from Fig.\:\:\hyperref[Fig8]{9(b)} that $\left\vert \psi_{tt}\left(
x_{1},x_{2}\right)  \right\vert ^{2}$ is almost the same at $\Gamma x=0$ and
$\Gamma x=5$, which indicates the contribution from the bound state is negligible.
Under the condition of $\kappa/\Gamma=100$, the nonrecipoal effect almost vanishes.

\section{CONCLUTION}\label{V}

In conclusion, we studied the transport of few photons in a
one-dimensional waveguide chirally
coupled to a dissipative nonlinear cavity. We derived the
scattering amplitudes of few photons analytically and
proved that the diode effect of few
photons is highly dependent on the intrinsic dissipation
of the cavity. First we discuss the single photon case. When $\kappa =\pm \left( \gamma _{1} -\gamma _{2}
\right)$ and $\omega_{k}=\omega_{a}$, the system works as a
single-photon diode which blocks the right-going or left-going photons, respectively. This means dissipation is necessary for the single-photon diode for chiral couplings  $\gamma _{1} \neq\gamma _{2}$.   When $\kappa=\Gamma$
and $\gamma_{1}/\Gamma=0$ ($\gamma_{1}/\Gamma=1$), the system acts as a
perfect diode because the transmittance is unity form one side and zero from the other side. However, when $\kappa
\ll\Gamma$ and $\omega_{k}=\omega_{a}$, the nonlinear cavity can be regarded
as a perfect cavity, and there is no nonreciprocal effect.
When $\gamma_{1}=\gamma_{2}$, the single photon is completely reflected. When $\Gamma\ll\kappa
$, the cavity and the waveguide are weakly coupled, allowing
photons to propagate freely in the waveguide.

There are also a similar diode effects in the
two-photon scattering, but at specific position $x$. We
analytically solved the equation $\psi_{tt}\left(  x_{1},x_{2}\right)  =0$ and found the relation
between $\left\vert x\right\vert $ and system parameters. When $\kappa\ll\Gamma$, we
found that the nonreciprocal effect is more pronounced than when $\kappa\approx\Gamma$ at the bound state region \cite{36}. The system exhibits reciprocity in regions far from the center of the bound states. When
$\kappa\gg\Gamma$, the transmission probability is almost the same at position $|x|$, because the bound state part almost vanishes due to weak coupling $\Gamma$. In summary, a
optical diode at few photon levels is realized by varying the ratio of $\kappa/\Gamma$. The working area and properties of the diode, along with the required parameters are found, which may find its applications in quantum networks.

\section*{Acknowledgements}

The authors thank  Jin-Lei Tan for helpful discussions. This
work was supported by the Scientific Research Fund of Hunan Provincial Education Department (Grant No. 23A0135), Natural Science Foundation  of Hunan Province, China (Grants No. 2024JJ10045 and No. 2022JJ30556),
Guizhou Provincial Basic Research Program (Natural Sci
ence) (Grant No. ZK[2024]021). National Natural Science Foundation of China (Grant No.~11704320). 

\begin{widetext}
\section*{Appendix:Two-photon scattering}

In this appendix we deduce the scattering of two left-incident photons in a
waveguide. We will convert the right-going mode into the even and odd modes by setting
$\hat{r}^{\dag}\left(  x\right)  =\left[  \sqrt{\gamma_{1}}\hat{c}_{e}^{\dag
}\left(  x\right)  +\sqrt{\gamma_{2}}\hat{c}_{o}^{\dag}\left(  x\right)
\right]  /\sqrt{\Gamma}.$ The incident state Eq. (\ref{14}) can be rewritten as
\begin{align}
\left\vert \psi_{in}\right\rangle  &  =\frac{\gamma_{1}}{\Gamma}\int\int
dx_{1}dx_{2}\varphi_{k}\left(  x_{1,}x_{2}\right)  \frac{1}{\sqrt{2}}\hat
{c}_{e}^{\dag}\left(  x_{1}\right)  \hat{c}_{e}^{\dag}\left(  x_{2}\right)
\left\vert \oslash\right\rangle +\frac{\gamma_{2}}{\Gamma}\int\int
dx_{1}dx_{2}\varphi_{k}\left(  x_{1,}x_{2}\right)  \frac{1}{\sqrt{2}}\hat
{c}_{o}^{\dag}\left(  x_{1}\right)  \hat{c}_{o}^{\dag}\left(  x_{2}\right)
\left\vert \oslash\right\rangle \nonumber\\
&  +\frac{\sqrt{2\gamma_{1}\gamma_{2}}}{\Gamma}\int\int dx_{1}dx_{2}%
\varphi_{k}\left(  x_{1,}x_{2}\right)  \hat{c}_{e}^{\dag}\left(  x_{1}\right)
\hat{c}_{o}^{\dag}\left(  x_{2}\right)  \left\vert \oslash\right\rangle ,
\tag{A1}%
\end{align}
We will introduce the scattering states of two photons in the odd
and even mode spaces.

The general two-photon scattering state of the system in the odd and even mode
spaces can be expressed as follows:
\begin{equation}
\left\vert \psi\right\rangle =\frac{\gamma_{1}}{\Gamma}\left\vert \psi
_{ee}\right\rangle +\frac{\gamma_{2}}{\Gamma}\left\vert \psi_{oo}\right\rangle
+\sqrt{\frac{\gamma_{1}\gamma_{2}}{2\Gamma^{2}}}\left\vert \psi_{oe}%
\right\rangle +\sqrt{\frac{\gamma_{1}\gamma_{2}}{2\Gamma^{2}}}\left\vert
\psi_{eo}\right\rangle , \tag{A2}%
\end{equation}
with%
\begin{subequations}
\begin{equation}
\left\vert \psi_{ee}\right\rangle =\int\int dx_{1}dx_{2}\varphi_{ee}\left(
x_{1},x_{2}\right)  \frac{1}{\sqrt{2}}\hat{c}_{e}^{\dag}\left(  x_{1}\right)
\hat{c}_{e}^{\dag}\left(  x_{2}\right)  \left\vert \varnothing\right\rangle
+\int dx\varphi_{ae}\left(  x\right)  \hat{c}_{e}^{\dag}\left(  x\right)
\hat{a}^{\dag}\left\vert \varnothing\right\rangle +\varphi_{aa}\frac{1}%
{\sqrt{2}}\hat{a}^{\dag}\hat{a}^{\dag}\left\vert \varnothing\right\rangle ,
\tag{A3}%
\end{equation}

\end{subequations}
\begin{equation}
\left\vert \psi_{oe}\right\rangle =\int\int dx_{1}dx_{2}\varphi_{oe}\left(
x_{1},x_{2}\right)  \hat{c}_{o}^{\dag}\left(  x_{1}\right)  \hat{c}_{e}^{\dag
}\left(  x_{2}\right)  \left\vert \varnothing\right\rangle +\int
dx\varphi_{oa}\left(  x\right)  \hat{c}_{o}^{\dag}\left(  x\right)  \hat
{a}^{\dag}\left\vert \varnothing\right\rangle , \tag{A4}%
\end{equation}

\begin{equation}
\left\vert \psi_{eo}\right\rangle =\int\int dx_{1}dx_{2}\varphi_{eo}\left(
x_{1},x_{2}\right)  \hat{c}_{e}^{\dag}\left(  x_{1}\right)  \hat{c}_{o}^{\dag
}\left(  x_{2}\right)  \left\vert \varnothing\right\rangle +\int
dx\varphi_{oa}\left(  x\right)  \hat{c}_{o}^{\dag}\left(  x\right)  \hat
{a}^{\dag}\left\vert \varnothing\right\rangle , \tag{A5}%
\end{equation}

\begin{equation}
\left\vert \psi_{oo}\right\rangle =\int\int dx_{1}dx_{2}\varphi_{oo}\left(
x_{1},x_{2}\right)  \frac{1}{\sqrt{2}}\hat{c}_{o}^{\dag}\left(  x_{1}\right)
\hat{c}_{o}^{\dag}\left(  x_{2}\right)  \left\vert \varnothing\right\rangle ,
\tag{A6}%
\end{equation}
where $\varphi_{ij}\left(  i,j=e,o,a\right)  $  is the amplitude of
the two photons; Subscript $i$, $j$ stand for one photon in mode $i$ and
the other in mode $j$; Subscript $a$ stands for the cavity mode. To satisfy
the exchange symmetry of photons, the amplitudes satisfy the relations:
$\varphi_{ee}\left(  x_{1},x_{2}\right)  =\varphi_{ee}\left(  x_{2}%
,x_{1}\right)  $, $\varphi_{oo}\left(  x_{1},x_{2}\right)  =\varphi
_{oo}\left(  x_{2},x_{1}\right)  $, $\varphi_{eo}\left(  x_{1},x_{2}\right)
=\varphi_{oe}\left(  x_{2},x_{1}\right)  $ and $\varphi_{ao}\left(  x\right)
=\varphi_{oa}\left(  x\right)  $.

In this paper, we focus on the two-photon transport of the frequency
$\omega=\omega_{k_{1}}+\omega_{k_{2}}$. Based on the time-independent
Schr\"{o}dinger equation, $\hat{H}\left\vert \psi\right\rangle $ $=$
$\omega\left\vert \psi\right\rangle $, we obtain a system of equations
for the scattering amplitudes:%
\begin{equation}
\left(  -i\upsilon_{c}\frac{\partial}{\partial x_{1}}-i\upsilon_{c}%
\frac{\partial}{\partial x_{2}}-\omega\right)  \varphi_{ee}\left(  x_{1}%
,x_{2}\right)  +\sqrt{\frac{\upsilon_{c}\Gamma}{2}}\delta\left(  x_{1}\right)
\varphi_{ae}\left(  x_{2}\right)  +\sqrt{\frac{\upsilon_{c}\Gamma}{2}}%
\delta\left(  x_{2}\right)  \varphi_{ae}\left(  x_{1}\right)  =0, \tag{A7}%
\end{equation}%
\begin{equation}
\left(  -i\nu_{c}\frac{\partial}{\partial x}+\omega_{a}-\omega-i\frac{\kappa
}{2}\right)  \varphi_{ae}\left(  x\right)  +\sqrt{2\upsilon_{c}\Gamma}%
\delta\left(  x\right)  \varphi_{aa}+\sqrt{\frac{\upsilon_{c}\Gamma}{2}%
}\varphi_{ee}\left(  0,x\right)  +\sqrt{\frac{\upsilon_{c}\Gamma}{2}}%
\varphi_{ee}\left(  x,0\right)  =0, \tag{A8}%
\end{equation}%
\begin{equation}
\left(  2\omega_{a}-\omega+2U-i\kappa\right)  \varphi_{aa}+\sqrt{2\upsilon
_{c}\Gamma}\varphi_{ae}\left(  0\right)  =0, \tag{A9}%
\end{equation}%
\begin{equation}
\left(  -i\upsilon_{c}\frac{\partial}{\partial x_{1}}-i\upsilon_{c}%
\frac{\partial}{\partial x_{2}}-\omega\right)  \varphi_{oe}\left(  x_{1}%
,x_{2}\right)  +\sqrt{\upsilon_{c}\Gamma}\delta\left(  x_{2}\right)
\varphi_{oa}\left(  x_{1}\right)  =0, \tag{A10}%
\end{equation}%
\begin{equation}
\left(  -i\upsilon_{c}\frac{\partial}{\partial x}+\omega_{a}-\omega
-i\frac{\kappa}{2}\right)  \varphi_{oa}\left(  x\right)  +\sqrt{\upsilon
_{c}\Gamma}\varphi_{oe}\left(  x,0\right)  =0, \tag{A11}%
\end{equation}%
\begin{equation}
\left(  -i\upsilon_{c}\frac{\partial}{\partial x_{1}}-i\upsilon_{c}%
\frac{\partial}{\partial x_{2}}-\omega\right)  \varphi_{oo}\left(  x_{1}%
,x_{2}\right)  =0. \tag{A12}%
\end{equation}
We use $\varphi_{ee}\left(  x,0\right)  =\left[  \varphi_{ee}\left(
x,0^{+}\right)  +\varphi_{ee}\left(  x,0^{-}\right)  \right]  /2$,
$\varphi_{ae}\left(  0\right)  =\left[  \varphi_{ae}\left(  0^{+}\right)
+\varphi_{ee}\left(  0^{-}\right)  \right]  /2$, $\varphi_{oe}\left(
x,0\right)  =\left[  \varphi_{oe}\left(  x,0^{+}\right)  +\varphi_{ee}\left(
x,0^{-}\right)  \right]  /2$ for the discontinuous points. Solving the
equations with the incoming state conditions and the discontinuity relations,%

\begin{equation}
\varphi_{ee}\left(  0^{+},x\right)  =\varphi_{ee}\left(  0^{-},x\right)
-i\sqrt{\frac{\Gamma}{2\upsilon_{c}}}\varphi_{ae}\left(  x\right)  ,\tag{A13}%
\end{equation}%
\begin{equation}
\varphi_{ee}\left(  x,0^{+}\right)  =\varphi_{ee}\left(  x,0^{-}\right)
-i\sqrt{\frac{\Gamma}{2\upsilon_{c}}}\varphi_{ae}\left(  x\right)  ,\tag{A14}%
\end{equation}%
\begin{equation}
\varphi_{oe}\left(  x_{1},0^{+}\right)  =\varphi_{oe}\left(  x_{1}%
,0^{-}\right)  -i\sqrt{\frac{\Gamma}{\upsilon_{c}}}\varphi_{oa}\left(
x_{1}\right)  ,\tag{A15}%
\end{equation}%
\begin{equation}
\varphi_{ae}\left(  0^{+}\right)  =\varphi_{ae}\left(  0^{-}\right)
-i\sqrt{\frac{\Gamma}{\upsilon_{c}}}\varphi_{aa},\tag{A16}%
\end{equation}%
\begin{equation}
\varphi_{oe}\left(  0^{+},x_{2}\right)  =\varphi_{oe}\left(  0^{-}%
,x_{2}\right)  ,\tag{A17}%
\end{equation}%
\begin{equation}
\varphi_{oa}\left(  0^{+}\right)  =\varphi_{oa}\left(  0^{-}\right)
,\tag{A18}%
\end{equation}
we obtained the amplitude for the scattering state,
\begin{equation}
\varphi_{ee}\left(  x_{1},x_{2}\right)  =\frac{1}{\sqrt{2}}\left[
\varphi_{e,k_{1}}\left(  x_{1}\right)  \varphi_{e,k_{2}}\left(  x_{2}\right)
+\varphi_{e,k_{1}}\left(  x_{2}\right)  \varphi_{e,k_{2}}\left(  x_{1}\right)
\right]  +\left[  \theta\left(  x_{2}-x_{1}\right)  \theta\left(
x_{2}\right)  De^{i\omega\frac{x_{c}}{\upsilon_{c}}}e^{\left[  i\left(
\omega-2\omega_{a}\right)  -\left(  \kappa+\Gamma\right)  \right]  \frac
{x}{2\upsilon_{c}}}+\left(  x_{2}\leftrightarrow x_{1}\right)  \right]
,\tag{A19}%
\end{equation}%
\begin{equation}
\varphi_{ae}\left(  x_{i}\right)  =\theta\left(  -x_{i}\right)  \left[
m_{a1}e^{ik_{1}x_{i}}+m_{a2}e^{ik_{2}x_{i}}\right]  +\theta\left(
x_{i}\right)  \left[  \beta_{k_{1}}e^{ik_{1}x_{i}}+\beta_{k_{2}}e^{ik_{2}%
x_{i}}+\chi e^{i\frac{\rho}{v_{c}}x_{i}}\right]  ,\tag{A20}%
\end{equation}%
\begin{equation}
\varphi_{oe}\left(  x_{1},x_{2}\right)  =\frac{1}{\sqrt{2}}\left[
\varphi_{o,k_{1}}\left(  x_{1}\right)  \varphi_{e,k_{2}}\left(  x_{2}\right)
+\varphi_{e,k_{1}}\left(  x_{2}\right)  \varphi_{o,k_{2}}\left(  x_{1}\right)
\right]  ,\tag{A21}%
\end{equation}%
\begin{equation}
\varphi_{oo}\left(  x_{1},x_{2}\right)  =\frac{1}{\sqrt{2}}\left[
\varphi_{o,k_{1}}\left(  x_{1}\right)  \varphi_{o,k_{2}}\left(  x_{2}\right)
+\varphi_{o,k_{1}}\left(  x_{2}\right)  \varphi_{o,k_{2}}\left(  x_{1}\right)
\right]  ,\tag{A22}%
\end{equation}%
\begin{equation}
\varphi_{oa}\left(  x_{i}\right)  =\varsigma_{k_{1}}e^{ik_{1}x_{i}}%
+\varsigma_{k_{2}}e^{ik_{2}x_{i}},\tag{A23}%
\end{equation}%
\begin{equation}
\varphi_{aa}=-\frac{1}{\sqrt{2}}\frac{\sqrt{\upsilon_{c}\Gamma}\left(
u_{a1}+u_{a2}\right)  }{\left(  \omega_{a}-\frac{\omega_{k_{1}}+\omega_{k_{2}%
}}{2}\right)  +U-i\frac{\kappa+\Gamma}{2}},\tag{A24}%
\end{equation}%
\begin{equation}
\varphi_{e,k_{i}}\left(  x_{j}\right)  =\frac{1}{\sqrt{2\pi}}\left[
\theta\left(  -x_{j}\right)  +t_{k_{i}}\theta\left(  x_{j}\right)  \right]
e^{ik_{i}x_{j}},\tag{A25}%
\end{equation}%
\begin{equation}
\varphi_{o,k_{i}}\left(  x_{j}\right)  =\frac{1}{\sqrt{2\pi}}e^{ik_{i}x_{j}%
},\tag{A26}%
\end{equation}
where $\theta\left(  x\right)  $ is the step function; $\omega_{k_{i}}$ (for
$i$ $=$ $1,2$) is the frequency of a single photon; $x_{c}$ $=$ $\left(
x_{1}+x_{2}\right)  /2$ and $x=x_{2}-x_{1}$ are the center-of-mass and the
relative coordinates, respectively;%
\begin{equation}
t_{k_{i}}=\frac{\omega_{k_{i}}-\omega_{a}+i\frac{\kappa-\Gamma}{2}}%
{\omega_{k_{i}}-\omega_{a}+i\frac{\kappa+\Gamma}{2}},\tag{A27}%
\end{equation}%
\begin{equation}
D=\frac{\sqrt{\Gamma}}{i\sqrt{2\upsilon_{c}}}\chi,\tag{A28}%
\end{equation}%
\begin{equation}
m_{a1}=\frac{\sqrt{\upsilon_{c}\Gamma}}{2\pi}\frac{1}{\omega_{k_{2}}%
-\omega_{a}+i\frac{\kappa+\Gamma}{2}},\tag{A29}%
\end{equation}%
\begin{equation}
m_{a2}=\frac{\sqrt{\upsilon_{c}\Gamma}}{2\pi}\frac{1}{\omega_{k_{1}}%
-\omega_{a}+i\frac{\kappa+\Gamma}{2}},\tag{A30}%
\end{equation}%
\begin{equation}
\chi=\frac{i4\pi\sqrt{\Gamma}U}{\sqrt{\upsilon_{c}}\left(  \omega_{a}%
-\frac{\omega_{k_{1}}+\omega_{k_{2}}}{2}+U-i\frac{\kappa+\Gamma}{2}\right)
}m_{a2}m_{a1},\tag{A31}%
\end{equation}%
\begin{equation}
\rho=\omega-\omega_{a}+i\frac{\kappa+\Gamma}{2},\tag{A32}%
\end{equation}
and $\beta_{k_{i}}$ $=$ $m_{ai}t_{k_{i}}$ and $\varsigma_{k_{i}}=$
$m_{ai}/\sqrt{2}$.
\end{widetext}

\end{document}